# SECURING DATA TRANSFER IN THE CLOUD THROUGH INTRODUCING IDENTIFICATION PACKET AND UDT -AUTHENTICATION OPTION FIELD: A CHARACTERIZATION

Danilo Valeros Bernardo[1] and Doan B Hoang[2]

[1]i-NEXT, Faculty of Engineering and Information Technology,
The University of Technology- Sydney
`bernardan@gmail.com`
[2] i-NEXT, Faculty of Engineering and Information Technology,
`dhoang@it.uts.edu.au`

## ABSTRACT

*The emergence of various technologies has since pushed researchers to develop new protocols that support high density data transmissions in Wide Area Networks. Many of these protocols are TCP protocol variants, which have demonstrated better performance in simulation and several limited network experiments but have limited practical applications because of implementation and installation difficulties. On the other hand, users who need to transfer bulk data (e.g., in grid/cloud computing) usually turn to application level solutions where these variants do not fair well. Among protocols considered in the application level solutions are UDP-based protocols, such as UDT (UDP-based Data Transport Protocol) for cloud /grid computing. Despite the promising development of protocols like UDT, what remains to be a major challenge that current and future network designers face is to achieve survivability and security of data and networks. Our previous research surveyed various security methodologies which led to the development of a framework for UDT. In this paper we present lower-level security by introducing an Identity Packet (IP) and Authentication Option (AO) for UDT.*

## KEYWORDS

*Next Generation Network Protocol, High Speed Bandwidth, UDT, AO, IP, SHA-1, SHA-256, MD5, Cloud, GRID*

## 1. INTRODUCTION

Many TCP protocol variants have demonstrated better performance in simulation and several limited network experiments [19]. However, practical use in real applications of these protocols is still very limited because of the implementation and installation difficulties. On the other hand, users who need to transfer bulk data (e.g., in grid/cloud computing) usually turn to application level solutions where these variants do not fair well. Among protocols considered in the application level solutions are UDP-based protocols, such as UDT (UDP-based Data Transport Protocol).

UDT is considered one of the most recently developed new transport protocols with congestion control algorithms. It was developed to support next generation high-speed networks, including wide area optical networks. It is considered a state-of-the-art protocol, which promptly addresses various infrastructure requirements for transmitting data in high-speed networks. Its development, however, creates new vulnerabilities because like many other protocols, it is designed to rely solely on existing security mechanisms for existing protocols such as the Transmission Control Protocol (TCP) and User Datagram Protocol (UDP). Some of these security mechanisms cannot be used to absolutely protect UDT, just as security mechanisms devised for wired networks cannot be used to protect unwired ones. Both the recently developed UDT and the decades-old TCP/UDP lack a well-thought-out security architecture that addresses problems in today's networks.

In this paper, we extend our previous works [7-13] and introduce additional approaches that can further assist network and security investigators, designers, and users who consider and incorporate security in the important layers when implementing UDT across wide area networks. These can support security architectural designs of UDP-based protocols as well as assist in the future development of other state-of-the-art fast data transfer protocols.

## 1.1 Background

The emergence of advanced high speed networks has created opportunities for new technologies to prosper. From mobile to wireless and network heterogeneous systems, issues of speed and performance may finally be addressed. However, due to the limitations of today's network protocols, it is difficult to scale data intensive applications from local clusters and metropolitan area networks to wide area networks. Many existing solutions are built around the existing constraints of existing mature protocols, such as TCP/IP.

Recent developments in network research introduced UDT, considered to be one of the next generation of high performance UDP- data transfer protocols [10]. UDT introduces a new three-layer protocol architecture that is composed of a connection flow multiplexer, enhanced congestion control, and resource management. The UDT design allows protocol to be shared by parallel connections and to be used by future connections. It improves congestion control and reduces connection set-up time.

It provides [8,9,10,11] better usability by supporting a variety of network environments and application scenarios [8,9,10,11]. It addresses TCP's limitations by reducing the overhead required to send and receive streams of data.

The choice of UDT in this paper is significant for several reasons: UDT as a newly designed next generation protocol is considered one of the most promising and fastest protocols ever created that operates either on top of the UDP or IP. It is a reliable UDP based application level data transport protocol for distributed data intensive applications over wide area and optical high-speed networks.

## 1.2 UDT Fast Data Transfer Protocol

UDT is a UDP-based approach [12] and is considered to be the only UDP-based protocol that employs a congestion control algorithm targeting shared networks. It is a new application level protocol with support for user-configurable control algorithms and more powerful APIs. It has no security mechanism to protect itself from adversaries [12,19].

### 1.2.1 Packet Structures

UDT is designed to have two packet structures: data packets and control packets. The packets are distinguished by the first bit (flag bit) of the packet header. The data packet header starts with 0, while the control packet header starts with 1 (fig. 1) [19].

**Data Packet**

| \0 or 1 | | Sequence Number | 31 bit |
|---|---|---|---|
| FF | 0 | Message Number | 29 bit |
| | | Time Stamp | 32 bit |

| 0 | Packet | Information | User defined |
|---|---|---|---|
| 1 | type | 15 | types |

| 1 | Type | | Extended Type | 31 bit |
|---|---|---|---|---|
| X | ACK Sub - Sequence Number | | | |
| | Time Stamp | | | |
| | Control Information | | | |

**Control Packet**

*Figure. 1: UDT packet header structures [19]. The first bit of the packet header is a flag indicating if this is a data or control packet. Data packets contain a 31-bit sequence number, 29-bit message number, and a 32-bit time stamp. Conversely, for control packet headers bits 1-15 are the packet type information and bits 16 -31 can be used for user-defined types. The detailed control information depends on the packet type.*

The packet sequence number uses 31 bits after the flag bit. It uses packet-based sequencing, which means the sequence number is increased by 1 for each data packet sent. The sequence number is wrapped after it is increased to the maximum number ($2^{31}$ -1) [7,12,19].

As in other protocols such as DCCP, the sequence number is used to arrange packets into sequence in order to detect loss [2,19] and network duplicates; it is also used to protect against attackers, half-open connections, and delivery of very old packets. Every packet carries a sequence number. Most packet types include an acknowledgment number, which is carried in a control packet – the second packet structure of UDT. The control packet is parsed according to the structure if the flag bit of a UDT packet is 1.

UDT sequence numbers are data packet based. Like TCP, they are generated by each endpoint increased by 1. Both data and control packets that do not carry user data increment the sequence number. UDT is not a true reliable protocol as there are no real retransmissions. Even if there were, retransmissions of data packets would also predictably increment the sequence number [2]. However, this allows UDT implementations to include features such as detecting network duplications, retransmissions, and loss.

Another characteristic of UDT is that it is a connection-oriented duplex protocol that supports data streaming and partial reliable messaging. It also uses rate-based congestion control (rate control) and window-based flow control to regulate outgoing traffic. This was designed so that the rate control updates the packet-sending period at constant intervals, whereas the flow control updates the flow window size each time an acknowledgment packet is received. This protocol has been expanded to satisfy further requirements for both network research and applications development. This expansion is called Composable UDT and was designed to complement the kernel space network stacks. However, this feature is intended for the following:

- Implementation and deployment of new control algorithms;
- Data transfer through private links that are implemented using Composable UDT;
- Supporting application-aware algorithms through Composable UDT; and
- Ease of testing new algorithms for kernel space when using Composable UDT compared to modifying an OS kernel.

The Composable UDT library implements a standard TCP Congestion Control Algorithm (CTCP). CTCP can be redefined to implement more TCP variants, such as TCP (low-based) and TCP (delay-based). The designers [19] emphasised that the Composable UDT library does not implement the same mechanisms in the TCP specification. TCP uses byte-based sequencing, whereas UDT uses packet-based sequencing. The designers stressed that this does not prevent CTCP from simulating TCP's congestion avoidance behaviour [7-13].

UDT was designed with the Configurable Congestion Control (CCC) interface, which is composed of four categories. The designers [19] included: 1) control event handler callbacks; 2) protocol behaviour configuration; 3) packet extension; and 4) performance monitoring.

As shown in Table 1 (Appendix), UDT's services/features can be used for bulk data transfer and streaming data processing. TCP cannot be used for this type of processing because it has two problems. First, the link must be clean (little packet loss) for it to fully utilize the bandwidth. Second, when two TCP streams start at the same time, a stream will be starved due to the RTT bias problem; thus, the data analysis process will have to wait for the slower data stream.

Moreover, UDT can offer streaming video to many clients. It can also provide selective streaming for each client when required, but its data reliability control needs to be handled by the application since UDP, which it operates on, cannot send data at a fixed rate.

### 1.2.2 UDT Implementation

One example of the implementation of UDT is the Sloan Digital Sky Survey (SDSS) project [40,44], which is mapping in detail one quarter of the entire sky, determining the positions and brightness of more than 300 million celestial objects. It measures distances to more than a million galaxies and quasars. The data from the SDSS project so far has increased to 13 terabytes and continues to grow and transmitted across Australia, Japan, South Korea, and China [19].

Securing data during its operations across network layers is therefore imperative in ensuring UDT itself is protected when implemented. The challenge to achieve  confidentiality, data integrity, and privacy for communication links and the challenge to address the complexity of running streaming applications over the Internet and through wireless and mobile devices, continue to mount.

In our previous works we presented a framework which adequately addresses vulnerability issues by implementing security mechanisms in UDT while maintaining transparency in data delivery. The development of the framework was based on the analyses drawn from the source codes of UDT found at SourceForge.net. The source codes were analyzed and tested on Win32 and Linux environments to gain a better understanding on the functions and characteristics of this new protocol.

Network and security simulations tools such as NS2 [31], EMIST [32], and proprietarily developed vulnerability testing tools were used to perform various simulation and experiments. Most of the security vulnerability tests, however, were conducted through simple penetration and traffic load tests. Tests were also performed on simulated environment running IPSec and developed simple solutions using Kerberos to test the method.

The results found in [10,11] provided significant groundwork in the development of a methodology which composes a variety of mechanisms to secure UDT against various adversaries, such as Sybil, addresses, man-in-the-middle and the most common, DoS attacks.

### 1.2.3   Unsecured UDT

UDT is at the application layer or on top of UDP. Subsequently data that it carries are required to be transmitted securely and correctly by each application.

UDT implementations are based on generic libraries [19] and through the its five application-dependent components, such as the API module, the sender, receiver, and UDP channel, as well as four data components: sender's protocol buffer, receiver's protocol buffer, sender's loss list and receiver's loss list, require that UDT security features must be implemented on an application basis [9,10,12].

The contention, therefore, for the need of security mechanisms of the new UDT is moreover derived from four important observations [2,7-13,19].

- Absence of an inherent security mechanism, such as checksum for UDT
- Dependencies on user preferences and implementation on the layer on which it is implemented
- Dependencies on existing security mechanisms of other layers on the stack
- Dependencies on TCP/UDP which are dependent on nodes and their addresses for high speed data transfer protocol leading to a number of attacks such as neighborhood, Sybil and DoS (Denial of Service) attacks

We contend the need to minimize its sending rates [7-11] in retransmissions, and to introduce its own checksum and Identity Packet (IP), and Authentication Option (AO) in its design.

This paper support the importance of implementing security in UDT, however, the introduction of other security mechanisms to secure UDT is presented to address its vulnerabilities to adversaries exploiting the application, transport, and IP layers.

### 1.3 Related Works

We  [8-13] presented a security framework highlighting the need to secure UDT and its existing features in comparison to TCP and UDP (Appendix table 1). The work focuses on UDT's position in the layer architecture which provides a layer-to-layer approach to addressing security (see figures 1 and 2). Its implementation relies on proven security mechanisms developed and implemented on existing mature protocols.

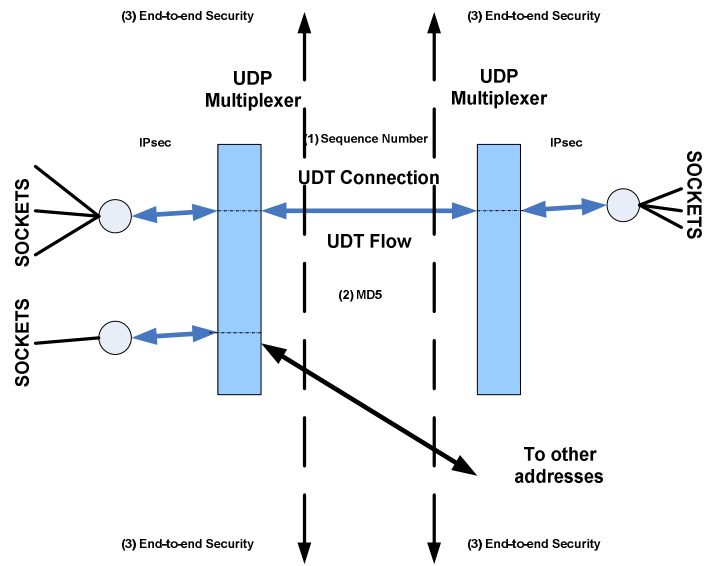

Figure 1: UDT in Layer Architecture. UDT is in the application layer above UDP. The application exchanges its data through UDT socket, which then uses UDP socket to send or receive data [7-13].

A summary of these security mechanisms and their implementations are presented in fig. 2.

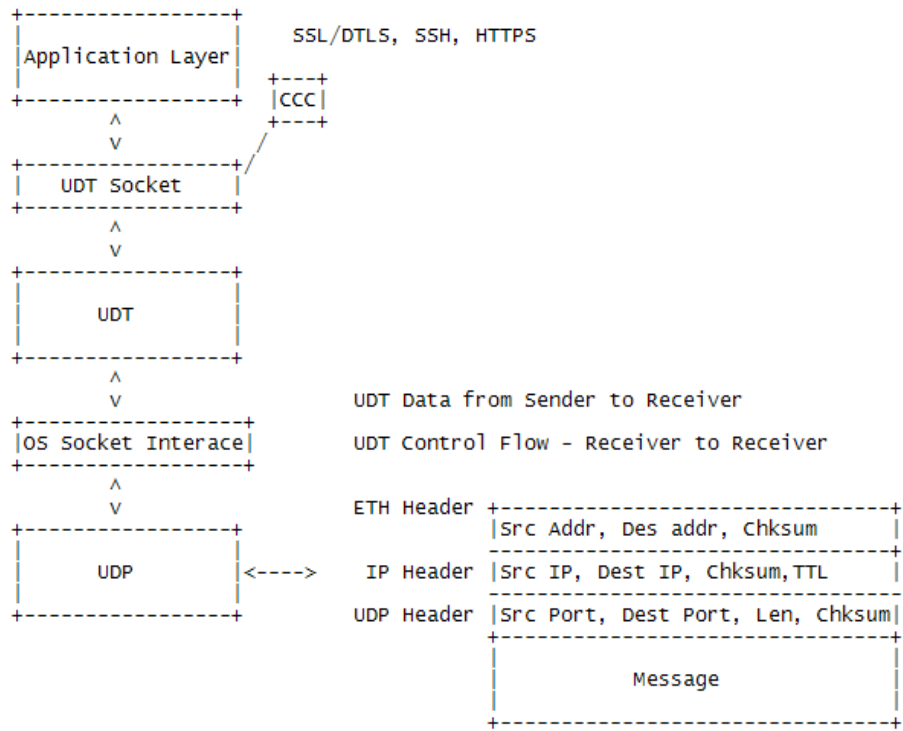

Figure 2. : UDT in Layer Architecture. UDT is in the application layer above UDP. The application exchanges its data through UDT socket, which then uses UDP socket to send or receive data .

### 1.4 Contributions

Past investigations [10,12] yield a developed practical and unified approach to securing UDT. The results make our approach attainable for other future protocols. We have used existing but novel methods for UDT and conducted experimentally-validated characterisations of UDT behaviours and its Socket API.

The proposed approaches are useful in several ways. They are presented to support future protocol implementations, and provided as a basis for design of both lower and higher-level communication layers and in the development of new protocols. To the best of our knowledge our contributions to UDT security is first in the literature.

## 2. METHODOLOGY

A comprehensive overview of the basic security mechanisms [7,8,9,12] for UDT was presented and as the research progressed, the introduction of additional set of methods in securing UDT was presented, ranging from securing UDT from the IP/Network Layers to the Transport and Application Layers.

The following methods are introduced. Firstly, utilizing IPsec (RFC 2401); however, for a number of reasons, this is only suitable for some applications [3,18]. Secondly, designing a custom security mechanism on the application layer using API, such as GSS-API [12,43], or a custom security mechanism on IP layer, such as HIP-CGA [1,21,23,38], or SASL[28]. Thirdly, integrating DTLS [16,35] on the transport layer, and lastly, which this paper proposes is the introduction of UDT Identity Packet and Authentication Option –AO.

The following are so far the approaches presented that are significant for application and transport layer- based authentication and end-to-end security for UDT. They are as follows:

### 2.1 IP layer

- Host Identity Protocol (HIP) [38]
- Self-certifying addresses using CGA [1]
- HIP+CGA [1,21,23,38]
- IPsec – IP security [3]

### 2.2 Session/Application and Transport Layers

- GSS-API - Generic Security Service Application  Program Interface [1,21,23,38]
- UDT-AO – Authentication Option
- SASL - Simple Authentication and Security Layer (SASL) [28]
- DTLS – Data Transport Layer Security [16,35]

[7-13] presented an overview on securing UDT implementations in various layers. However securing UDT in application and other layers needs to be explored in future UDT deployments in various applications.

There are application and transport layer-based authentication and end-to-end [12] security options for UDT. It is advocated the use of GSS-API in UDT in the development of an application using TCP/UDP. The use of Host Identity Protocol (HIP), a state of the art protocol, combined with Cryptographically Generated Addresses (CGA) is explored to solve the problems of address-related attacks. We presented a comprehensive review for each approach in our previous works [7-13].

In this paper, two approaches are introduced: (1) Identity Packet and (2) Authentication Option for UDT.

# 3. APPROACHES

## 3.1 Identity Packet within UDT

It was proposed [21,23] that the first packet of interaction should carry identity information. In today's Internet, the first packet (e.g. the TCP SYN) carries no higher-level data that provides adequate sender's information. On the initial transmission, it only carries the source IP address (network layer) and an initial sequence number (transport layer). The high-level data can be only exchanged or transmitted only after the complete ACK has been instantiated. Consequently, the receivers will not be able to establish from whom the data is coming from without using additional overhead.

UDT like TCP, contains no data which can be used to identify a user (except such information is contained within the (unencrypted) data part of the packet). While, the source and destination ports (TCP/UDP) in cooperation with the IP address of the sender and receiver can identify both participating parties [14,15,39] in the lower level, it carries no higher-level data that can identify the source before an application processes the packets received.

Network protocols, such as UDT, has a Sequence number field that provides identification, however, the initial value is determined by the implementer how the initial sequence number is chosen (e.g., randomly). The same is true for instance for the Window field used for congestion control. As the congestion control has a key influence on the overall performance of the protocol, many attempts to optimize them are done e.g., by operating system manufactures.

The lack of identity at the lower-level (network layer) in the existing network protocols has made achieving network security difficult. Several protocols developed and implemented on top of transport and IP layers usually created overhead and network incompatibilities. The development of a new technology that includes identity directly into the packets while retaining backward compatibility to allow an identity based approach to network security is a challenge. This has been introduced through placing encrypted and digitally signed identity information into the packets by developing applications that become part of the stack that add digitally signed identity information to the packets and decode the information of any incoming connection attempts.

For UDT to be utilized in tomorrow's Internet it has to accommodate higher-level data association while maintaining its dependency on the low-level protocol such as TCP and UDP. In other works [14,15] the initial packet of any association, which is called the rendezvous packet [15], carries high-level information to initiate the association. This provides the receiving entity the information on which it can decide on whether to process the first packet of an association. This information can be delivered in a reliable manner , through cryptographically protecting it prior and during the transmission[15].

A mechanism for "first packet identity" (see figure 3) within UDT should be devised that is robust enough that a receiver cannot be flooded by requests that require to take excessive action before they have verified the identity and trust at the application level. This information can be created using user-defined types field + information. It is possible to delegate this decision-making based on the first packet identity to a guard machine that can take on the burden and risk of overload from suspect sources.

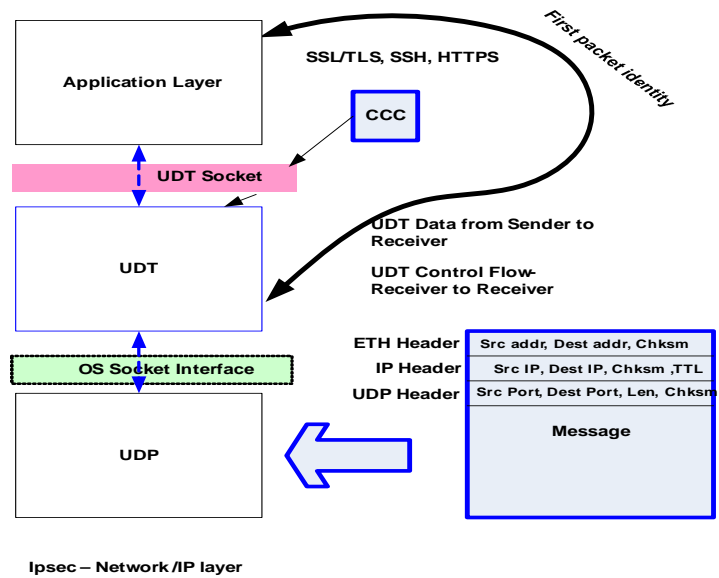

Figure 3. First packet Identity in UDT in Layer Architecture. The application exchanges its data through UDT socket, which then uses UDP socket to send or receive data through an encrypted mechanism.

## UDT Packet

| 0 1 | Packet type | Information 15 | User defined types | |
|---|---|---|---|---|
| 1 | Type | *IDENTITY packet* | Extended Type | 31 bit |
| X | | ACK Sub – Sequence Number | | |
| | | Time Stamp | | |
| | | Control Information | | |

UDT needs to build on the identity representation that is used at the application level, because that may not be visible to the routes (e.g., it might be encrypted see figure 4), it may reveal too many attributes of the user, and it may not be associated with each transmission unit (applications may be willing to install more state than routers are). It needs to be robust to deal with resource usage and flooding attacks.

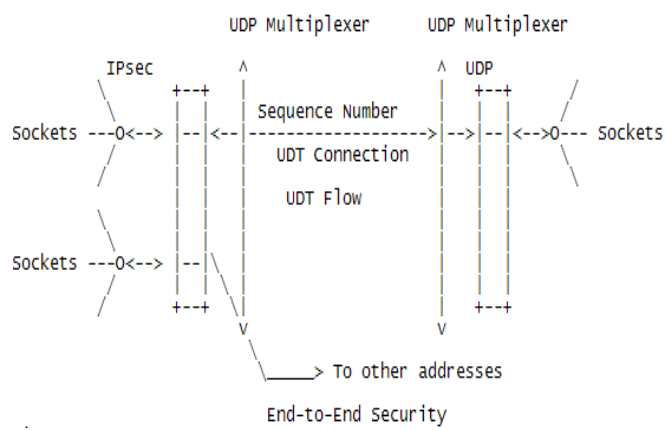

Figure 4. Conceptual design of UDT flow using end-to-end security. First packet identity is transmitted through IPsec[23].

## 3.2 UDT Option for Message Authentication

In this section ,we introduce a mechanism – UDT-Authentication Option (AO) to secure UDT - a new UDP-based data transport protocol. We describe this mechanism through the introduction of UDT extension to achieve security. We evaluate UDT-AO through the use of existing message authenticity for other protocols such as TCP. We review existing message protection that can act like a signature for UDT segment incorporating information known only to the connection end points. Since UDT is operating on UDP for high speed data transfer, we propose the creation of a new option in UDT that can significantly reduce the danger of attacks on applications running UDT. This can maintain message integrity during data transmissions on high speed networks.

We propose the creation of an UDT-AO option which will be added with the UDP checksum. UDT is a connection oriented protocol. As such, it needs to include an OPTION for authentication when it is used in data transmission. This is because its connections like TCP, are likely to be spoofed [41].

In TCP, this option is part of the options (0-44 bytes) that occupy space at the end of the TCP header. In utilising the option in TCP [18], this needs to be enabled in the socket. A few systems support this option, which is identified as TCP_MD5SIG option. Note: for the purpose of conceptual analysis, we use MD5 in this paper.

*int opt = 1;* Enabling this option
*setsockopt(sockfd, IPPROTO_TCP, TCP_MD5SIG, &opt, sizeof(opt));*

Additionally, the option is included in the checksum. The option may begin on any byte boundary, and the header must be padded with zeros to make the header length a multiple of 32 bits.

Better authentication options in TCP are in progress to address the collision issue in MD5 [25,34]. However in this paper, the primary motivation is to initially introduce an option to allow UDT to protect itself against the introduction of spoofed segments into the connection stream, regardless what authentication schemes required in a given number of bytes. This option, like in TCP, will be included in the UDP header.

UDT is in user space above the network transport layer of UDP, it is dependent on UDP where this option is not available, thus becomes a challenge in its implementation. UDT, however, provides transport functionalities to applications.

To spoof a connection of UDT, an attacker would not only have to guess UDT's sequence numbers, but would also have had to obtain the password included in the MD5 digest. This password never appears in the connection stream, and the actual form of the password is up to the application, according to RFC 2385. It could during the lifetime of a particular connection so long as this change was synchronised on both ends (although retransmission can become problematical in some implementations with changing passwords).

To utilize this option in UDT, similar to TCP this needs to be enabled in the socket. A few systems support this option, which can be identified as UDT_MD5SIG or in the case of using SHA, UDT_SHASIG option.

```
int opt = 1;  Enabling this option
setsockopt(sockfd, IPPROTO_UDT, UDT_MD5SIG, &opt, sizeof(opt));
```

Likewise, the option is included in the UDP checksum. However, there is no negotiation for the use of this option in a connection (also in TCP) rather it is purely a matter of site policy whether or not its connections use the option.

### 3.2.1 Syntax for UDT Option

The proposed option can be applied on type 2 of the UDT header. This field is reserved to define specific control packets in the Composable UDT framework.

Every segment sent on a UDT connection to be protected against spoofing will similarly contain the 16-byte MD5 digest produced by applying the MD5 algorithm to these items in the following similar order required for TCP :

1. UDP pseudo header (Source and Destination IP    addresses, port number, and segment length)

2. UDT header + UDP (Sequence number and timestamp), and assuming a UDP checksum zero

3. UDT control packet or segment data (if any)

4. Independently-specified key or password, known to both UDTs and presumably connection-specific and

5. Connection key

The UDT packet header and UDP pseudo-header are in network byte order. The nature of the key is deliberately left unspecified, but it must be known by both ends of the connection, similar with TCP. A particular UDT implementation will determine what the application may specify as the key.

To calculate UDP checksum a "pseudo header" is added to the UDP message header. This includes:

IP Source Address        4 bytes
IP Destination Address 4 bytes
Protocol                      2 bytes
UDP Length                2 bytes

The checksum is calculated over all the octets of the pseudo header, UDP header and data. If the data contains an odd number of octets a pad, zero octet is added to the end of data. The pseudo header and the pad are not transmitted with the packet.

Upon receiving a signed segment, the receiver must validate it by calculating its own digest from the same data (using its own key) and comparing the two digest. A failing comparison must result in the segment being dropped and must not produce any response back to the sender. Logging the failure is advisable [22].

Unlike other TCP extensions (e.g., the Window Scale option [RFC1323]), the absence of the option in the SYN,ACK segment must not cause the sender to disable its sending of signatures. This negotiation is typically done to prevent some TCP implementations from misbehaving upon receiving options in non-SYN segments. In UDT it is ACK2 (ACK of ACK).

This is not a problem for this option, according to Heffernan [22], as similarly since the SYN, ACK sent during connection negotiation will not be signed and will thus be ignored. This is the same with ACK2 for UDT. The connection will never be made, and non-SYN segments (which does not exist in UDP) with options will never be sent. More importantly, the sending of signatures must be under the complete control of the application, not at the mercy of the remote host not understanding the option.

The MD5 digest is always 16 bytes in length, and the option would appear in every segment of a connection [22,25,34,42].

### 3.2.2 Implications

#### 3.2.2.1 Hashing Algorithm

MD5 algorithm has been found to be vulnerable to collision search attacks, and is considered by some to be insufficiently strong for this type of application [22,25,34,42].

However, we specify the MD5 algorithm for this option as basis of our argument to include AO to UDT. Systems that use UDT have been deployed operationally, and there was no "algorithm type" field defined to allow an upgrade using the same option number.

This does not, therefore, prevent the deployment of another similar option which uses another hashing algorithm (like SHA-1, SHA-256). Moreover, should most implementations pad the 18 byte option as defined to 20 bytes anyway, it would be just as well to define a new option which contains an algorithm type field.

In addressing this implication, we recommend using a more secure message algorithm such as SHA-1 or SHA-256.

### 3.2.2.2. Key configuration

It should be noted that the key configuration mechanism of routers may restrict the possible keys that may be used between peers. It is strongly recommended that an implementation be able to support at minimum a key composed of a string of printable ASCII of 80 bytes or less, as this is current practice in TCP [34].

## 4. CONCLUSIONS

New applications vary in traffic and connection characteristics in various communication links. Most of them still use TCP for data transfer because of its reliability and stability. Despite being widely used, there are a number of serious security flaws inherent in TCP and UDP [2,3,4]. There have also been performance issues in the implementation of large networks that require high bandwidths.

On the other hand, the current and future Internet has no mechanism to constrain hosts that offer and use excess traffic more than the required bandwidths. Particularly, it contains no defense against DDoS (distributed denial of service) attacks. It does not include an architectural approach to protect its own elements from any attack [15]. Its existing link encryption may be sufficient and efficient for securing connectivity but not securing the Internet per se. The link encryption can also create network and application incompatibilities.

In this paper, we introduced 'first packet identity' which needs to be instituted but devised in such as way that it is robust enough that a receiver cannot be flooded by requests that require the receiver to take excessive action before they have verified the identity and trust at the application level.

We also introduced and proposed a state-of-the-art security mechanism for UDT and for its future implementations to various network topologies by adding a field for AO, authentication options. We demonstrated the use of MD5 [25], while we encourage the use of other hash functions, such as SHA-1 or SHA-256.

The preceding discussions presented in this paper focused on the conceptual low-level protection of the end node since UDT relies on TCP and UDP for data delivery. We proposed the inclusion of identity on its packet header (IP) and Authentication Option (AO) before the transmission is validated at the application level.

Future works include the implementation of the proposed framework [7-13] and the conceptual approach presented in this paper to device an Identity Packet for UDT and introduce Authentication Option to protect data transmission. The works will include evaluating other existing network protocols in order to adequately secure current and future high speed networks, including the Internet.

## ACKNOWLEDGEMENTS

Acknowledgement goes to Db2Powerhouse for its Science and Technology Initiative (STI), research grant #DB2P01021 and anonymous reviewers for their constructive comments that helped improve this work.

# **APPENDIX**

Table 1

| Services/Features | UDT | TCP | UDP |
|---|---|---|---|
| Connection-oriented | Yes | Yes | No |
| Full duplex | Yes | Yes | Yes |
| Reliable data transfer | Yes | Yes | No |
| Partial-reliable data transfer/message | Yes | No | No |
| Ordered data delivery | Yes | Yes | No |
| Unordered data delivery | - | No | Yes |
| Flow control | Yes | Yes | No |
| Congestion control | Yes | Yes | No |
| ECN capable | Yes | Yes | No |
| Selective ACKs | Yes | Optional | No |
| Preservation of message boundaries | Yes | No | Yes |
| Path MTU discovery | Yes | Yes | No |
| Application PDU fragmentation | Dependent | Yes | No |
| Multistreaming | Dependent | Yes | No |
| Multihoming | Dependent | No | No |
| Protection against SYN flooding attacks | No | No | n/a |
| Allows half-closed connections | Yes | Yes | n/a |
| Reachability check | Yes | Yes | No |
| Psuedo-header for checksum | No | Yes | Yes |
| Time wait state | Yes | 4-tuple | n/a |
| **SecurityMethods** | **UDT** | **TCP** | **UDP** |
| Checksum | Applicable | Yes | Yes |
| GSS-API | Applicable | Yes | No |
| SASL | Applicable | Yes | No |
| HIP | Applicable | Yes | No |
| UDT-AO | Applicable | TCP-AO | No |
| DTLS | Applicable | Yes | Applicable |
| IPSec | Applicable | Yes | Applicable |